# THERMAL TRANSIENT MULTISOURCE SIMULATION USING CUBIC SPLINE INTERPOLATION OF ZTH FUNCTIONS

*Dirk Schweitzer*

Infineon Technologies AG

**ABSTRACT**

This paper presents a very straightforward method to compute the transient thermal response to arbitrary power dissipation profiles in electronic devices with multiple heat sources. Using cubic spline interpolation of simulated or measured unit power step response curves (Zth-functions), additional errors due to model reduction can be avoided. No effort has to be spent on the generation of compact models. The simple analytic form of the interpolating splines can be exploited to evaluate the convolution integral of the Zth-functions with arbitrary power profiles at low computational costs. An implementation of the algorithm in a spreadsheet program (EXCEL) is demonstrated. The results are in very good agreement with temperature profiles computed by transient Finite Element simulation but can be obtained in a fraction of the time.

## 1. INTRODUCTION

The trend towards higher integration and higher power densities in semiconductor devices has also driven the needs of methods to predict the transient temperature of devices with multiple heat sources. Since the computational effort for direct thermal simulation of transient switching processes with detailed (e.g. finite element) models is huge, several methods of model reduction have evolved which drastically reduce computing time.

If material properties and boundary conditions are independent of temperature, a single heat source is thermally fully characterized by its unit power step response or Zth function $Z_{th}(t)$. I.e. the temperature response $T(t)$ to any power dissipation profile $P(t)$ can be computed by the convolution integral

$$T(t) = T_0 + \int_0^t \dot{Z}_{th}(t-\tau)\cdot P(\tau)\, d\tau, \qquad (1)$$

$T_0$ being the ambient temperature [1]. For real materials with temperature dependent density, specific heat, and thermal conductivity, equation (1) holds not exactly true, but the errors can be neglected as long as the temperature difference $T(t) - T_0$ is not too large. For $N$ heat sources the mutual heating has to be taken into account, extending eq. (1) to

$$T_i(t) = T_0 + \sum_{j=1}^{N} \int_0^t \dot{Z}_{th\,ij}(t-\tau)\cdot P_j(\tau)\, d\tau, \qquad (2)$$

for the temperature $T_i(t)$ of the i-th heat source [2]. $Z_{th\,ij}(t)$ is a transfer function, which describes the temperature rise at the location of source i, when only source j is heated. Again the superposition principle strictly speaking holds true only for temperature independent material properties but is a good approximation in most cases.

Based on these principles, any reduced model which correctly reproduces the thermal (transfer-) Zth curves can be used to simulate the temperature development at the locations of the heat sources of a device. In most cases thermally equivalent RC networks, which had been fitted to simulated or measured power step responses, have been used for this purpose [2] [3] [4]. However the process of model reduction introduces new errors into the simulation in addition to the errors inherent to the underlying Zth functions. An obvious idea is therefore to use directly the Zth values $Z_{th\,ij}(t_i)$ sampled at times $t_i$ and interpolate at intermediate points of time.

## 2. INTERPOLATION OF THE ZTH FUNCTIONS

In the following it is assumed that a thermal unit power step response $Z_{th}(t)$ has been sampled at $m$ points of time $t_i$ which are equidistant on a logarithmic timescale. The interpolation can either be done on a linear timescale, i.e. for points

$$\{(t_i, Z_{th\,i})\}, \quad i = 1\,..\,\mathrm{m} \qquad (3)$$

or on a logarithmic timescale, i.e. for points

$$\{(z_i, Z_{th\,i})\}, \quad i = 1..\,\mathrm{m} \quad \text{with } z_i = \ln t_i \qquad (4)$$

For a Foster RC network (figure 1), the exact value of the Zth function of its junction node can be calculated by the equation

$$Z_{th}(t) = \sum_{i=1}^{n} R_{th\,i}\left(1 - e^{-\frac{t}{R_{th\,i} C_{th\,i}}}\right) \qquad (5)$$

for any time $t$. Using eq. (5) as a model function, one can easily perform some numerical experiments to compare the deviations between interpolated and exact $Z_{th}$ curves for different interpolation methods.

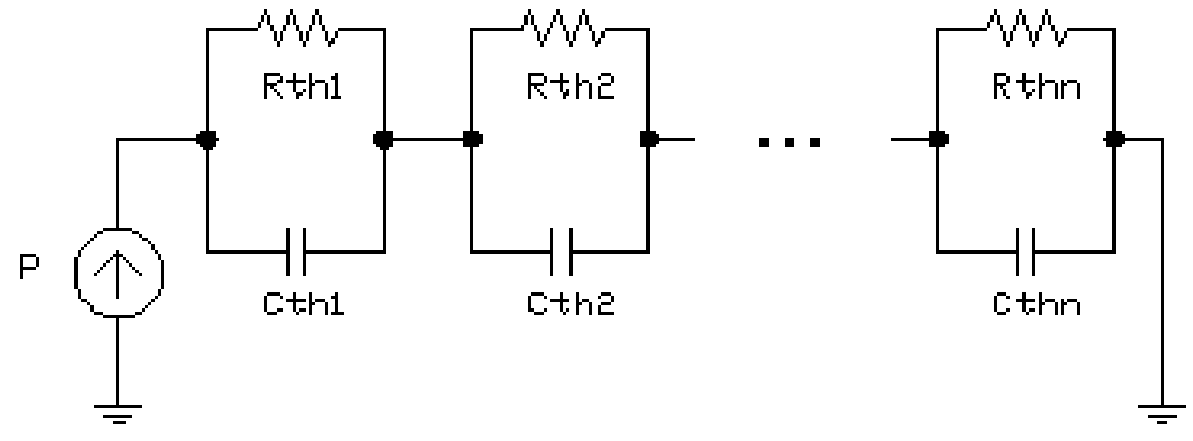

**Figure 1: Foster RC-network**

Infineon's TLE 6220 in a P-DSO-20 exposed pad package, an automotive device for engine control with four power dissipating channels on a single chip, will be used to demonstrate the methods presented herein. More details on the device model follow below. At the moment the RC parameters fitted for this device (table 1) just serve as an example. Using eq. (5), a data set containing *m*=51 points, equidistant on a logarithmic timescale, has been generated (figure 2).

| *i* | $R_{th\,i}$ [K/W] | $C_{th\,i}$ [Ws/K] |
|---|---|---|
| 1 | 0.07746 | 0.000273 |
| 2 | 0.48958 | 0.000365 |
| 3 | 1.55159 | 0.000772 |
| 4 | 2.70800 | 0.003253 |
| 5 | 1.45388 | 0.052680 |
| 6 | 2.51305 | 0.274200 |
| 7 | 1.05932 | 4.496140 |
| 8 | 0.11046 | 266.4490 |

**Table 1: Parameters of the test model**

Using cubic splines [5], this data set has been interpolated both on a linear timescale (interpolating between points given by eq. (3)) and on a logarithmic timescale (interpolating between points given by eq. (4)). The resulting relative error between interpolated and analytic thermal impedance,

$$rel.err = \frac{Z_{th}^{ip}(t) - Z_{th}(t)}{Z_{th}(t)}, \qquad (6)$$

is shown in figure 3. The maximum interpolation error is 0.03% for interpolation on a linear timescale and 0.01% for interpolation on a logarithmic timescale over the whole time range from 1.0e-6 s to 1000 s. These values depend of course on the shape of the interpolated Zth curve as well as on the distance of the interpolation points. But for a reasonable number of points, the interpolation error can be neglected. The errors that occur when fitting a RC-network to a unit power step response are typically in the range of a few percent (and often much larger for the transfer functions).

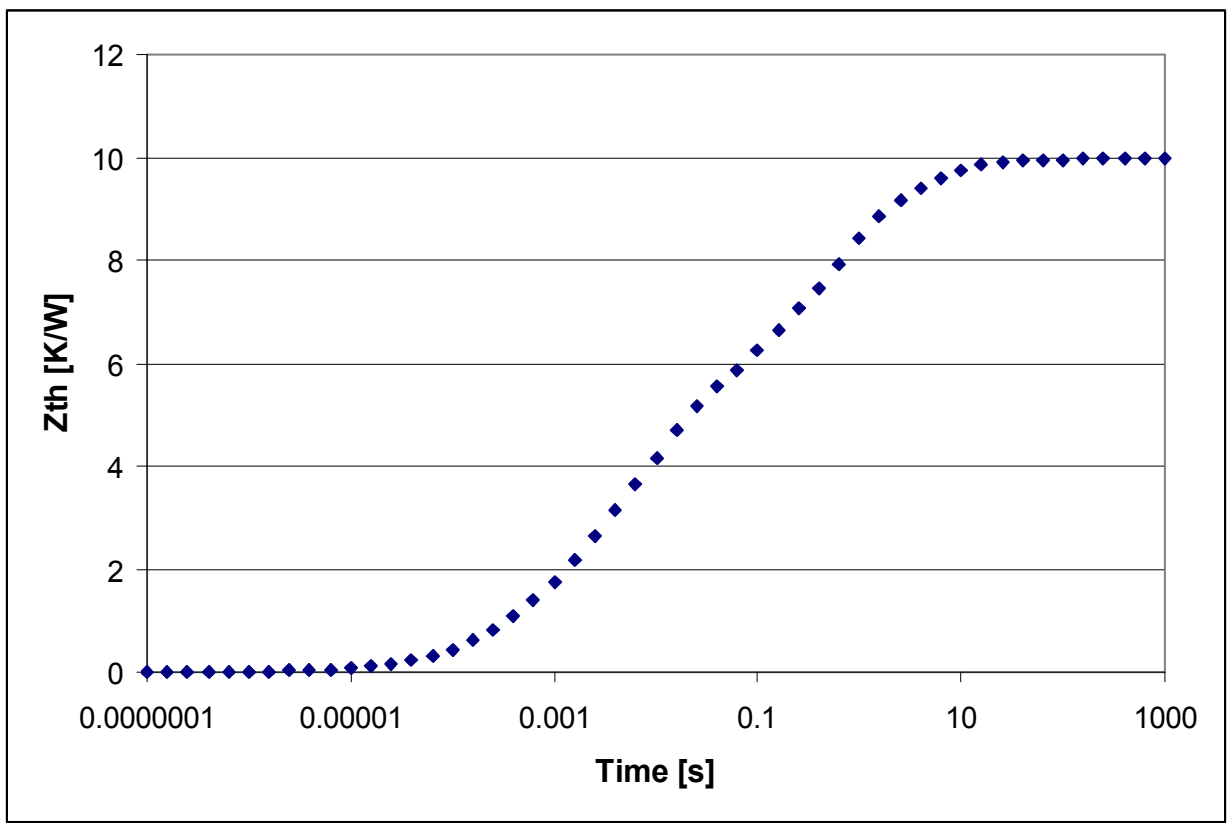

**Figure 2: Model curve to be interpolated**

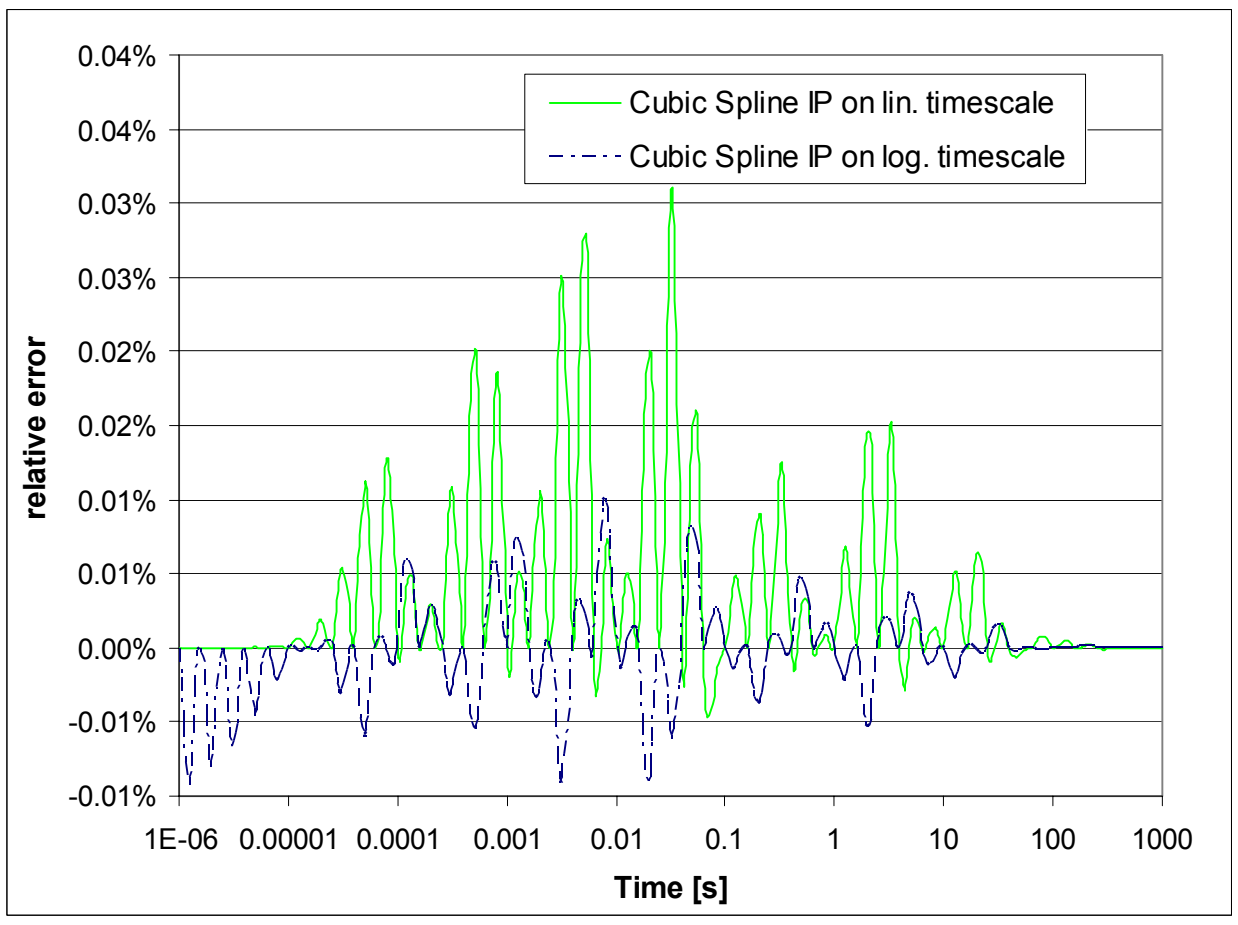

**Figure 3: Interpolation error (cubic splines)**

## 3. NUMERICAL CONVOLUTION

Cubic spline interpolation not only saves one from the trouble of fitting RC networks but can also be exploited to efficiently compute the convolution integrals in eq. (1) and (2). Any power dissipation profile *P(t)* can be approximated by a sequence of trapezoidal pulses as shown in figure 4. Using the superposition principle, the temperature response to the whole power profile is just the sum of the temperature response curves to all trapezoidal pulses.

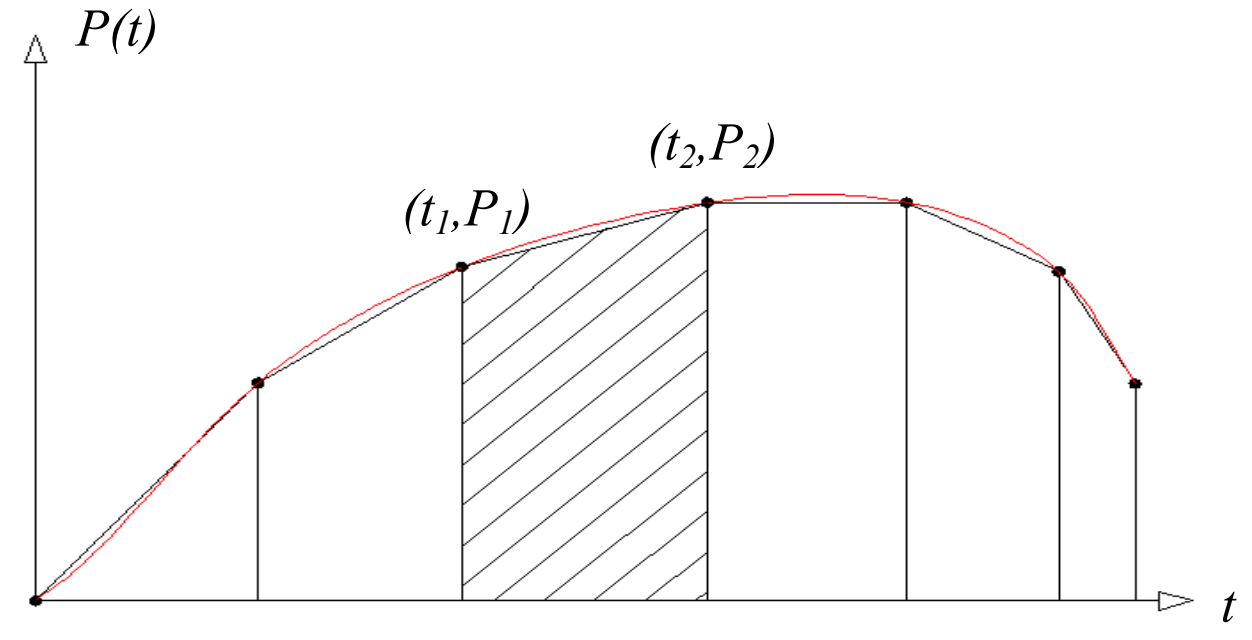

**Figure 4: Approximation of arbitrary power profiles**

For a single pulse, starting at time $t_1$ with power $P_1$ and ending at time $t_2$ with power $P_2$, partial integration of the convolution integral in eq.(1) yields:

$$\Delta T(t)=\begin{cases} 0, \quad t<t_1 \\ P_1 Z_{th}(t-t_1)+\frac{P_2-P_1}{t_2-t_1}\cdot\int_0^{t-t_1} Z_{th}(\tau)d\tau, \\ \qquad t_1\le t<t_2 \\ P_1 Z_{th}(t-t_1)-P_2 Z_{th}(t-t_2)+ \\ \frac{P_2-P_1}{t_2-t_1}\cdot\int_{t-t_2}^{t-t_1} Z_{th}(\tau)d\tau, \quad t\ge t_2 \end{cases} \quad (7)$$

with

$$\Delta T(t)=T(t)-T_0 .$$

The integrals in eq. (2) can be evaluated accordingly. I.e. to compute the temperature response *T(t)* for a trapezoidal pulse, not only the $Z_{th}(t)$ has to be evaluated, but also the integral:

$$IZ_{th}(t)\equiv\int_0^t Z_{th}(\tau)\,d\tau \quad (8)$$

Both can be easily computed since the (piecewise) interpolating splines are just third degree polynomials which can be integrated analytically:

$$Z_{th_i}^{ip}(t)=a_{i3}t^3+a_{i2}t^2+a_{i1}t+a_{i0}$$

$$\int_{t_i}^{t} Z_{th_i}^{ip}(\tau)\,d\tau=\tfrac{1}{4}a_{i3}(t-t_i)^4+\tfrac{1}{3}a_{i2}(t-t_i)^3+ \quad (9)$$
$$+\tfrac{1}{2}a_{i1}(t-t_i)^2+a_{i0}(t-t_i), \quad t_i\le t<t_{i+1}$$

with

$$Z_{th_i}^{ip}(t_i)=Z_{th_i} \quad \text{and} \quad Z_{th_i}^{ip}(t_{i+1})=Z_{th_{i+1}} .$$

Eq. (9) is the i-th interpolating spline between $Z_{th}$ values given at times $t_i$ and $t_{i+1}$ and its integral if the interpolation is done on a linear timescale. For cubic spline interpolation on a logarithmic timescale, the interpolating spline is

$$\widetilde{Z}_{th_i}^{ip}(z)=b_{i3}z^3+b_{i2}z^2+b_{i1}z+b_{i0}, \quad z=\ln t, \quad z_i\le z<z_{i+1} \quad (10)$$

with

$$\widetilde{Z}_{th_i}^{ip}(z_i)=Z_{th_i} \quad \text{and} \quad \widetilde{Z}_{th_i}^{ip}(z_{i+1})=Z_{th_{i+1}} .$$

The coefficients $a_{ij}$ ($b_{ij}$) are computed such as to ensure that the interpolating function is smooth in the first derivative and continuous in the second derivative (see [5] for a detailed description of the algorithm). In the latter case (eq. (10)) the transformation $t\rightarrow z$ yields

$$\int_{t_i}^{t} Z_{th_i}^{ip}(\tau)\,d\tau=\int_{t_i}^{t}\widetilde{Z}_{th_i}^{ip}(\ln\tau)\,d\tau=\int_{z_i}^{z}\widetilde{Z}_{th_i}^{ip}(\zeta)e^{\zeta}d\zeta \quad (11)$$

Which can be further evaluated using the elementary integrals

$$\int x\,e^x dx=e^x(x-1)$$
$$\int x^2\,e^x dx=e^x(x^2-2x+2) \quad (12)$$
$$\int x^3\,e^x dx=e^x(x^3-3x+6x-6).$$

to

$$\int_{z_i}^{z}\widetilde{Z}_{th_i}^{ip}(\zeta)e^{\zeta}d\zeta=e^z(c_{i3}z^3+c_{i2}z^2+c_{i1}z+c_{i0}) \quad (13)$$
$$-e^{z_1}(c_{i3}z_i^3+c_{i2}z_i^2+c_{i3}z_i+c_{i0})$$

with

$$\begin{aligned} c_{i3}&=b_{i3} \\ c_{i2}&=b_{i2}-3b_{i3} \\ c_{i1}&=b_{i1}-2b_{i2}+6b_{i3} \\ c_{i0}&=b_{i0}-b_{i1}+2b_{i2}-6b_{i3} \end{aligned} \quad (14)$$

The coefficients $a_{ij}$ or $b_{ij}$ and $c_{ij}$ should be stored for subsequent evaluations of $Z_{th}$ and $IZ_{th}$.

## 4. PERIODIC POWER PULSES

If the temperature response to a sequence of periodic power pulses is to be computed for a time $t$ much larger than the pulse period $t_p$ ($t >> t_p$), the contributions of all preceding pulses within the decay time $t_{decay}$ (i.e. the time it takes until the rise in temperature due to a single power pulse has dropped to a negligible value) must be taken into account. The ratio $t_{decay} / t_p$ and therefore the number of pulses to be included can be quite large, meaning a considerable computational effort.

But the difference between the temperature response to a periodic power excitation and to a constant power excitation with the corresponding average power vanishes quickly after the power has been switched off (figure 5). In the long run the decay curve depends merely on the total energy dissipated during the pulse(s), not on the specific shape of the power profile *P(t)*.

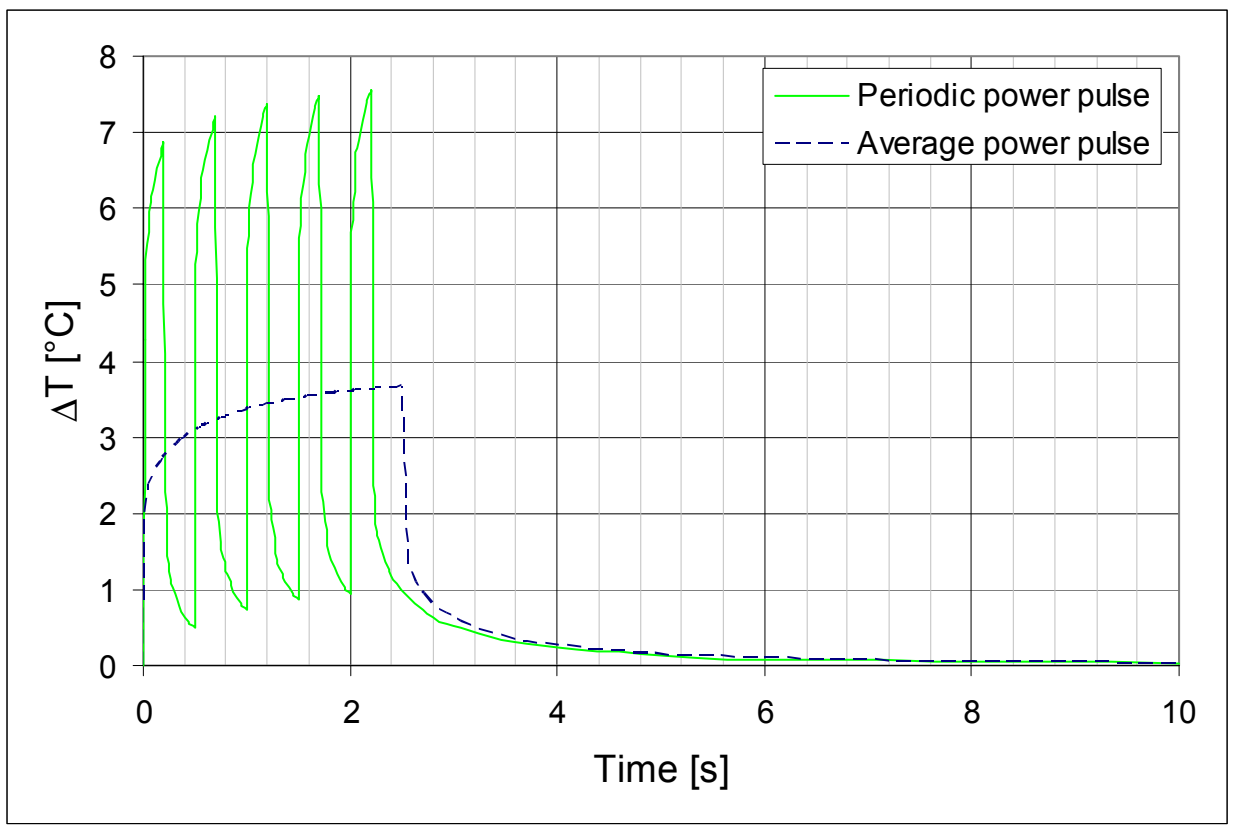

**Figure 5: Temperature response curves for a periodic sequence of 5 pulses and for the corresponding constant pulse with average power $P_{avg}$.**

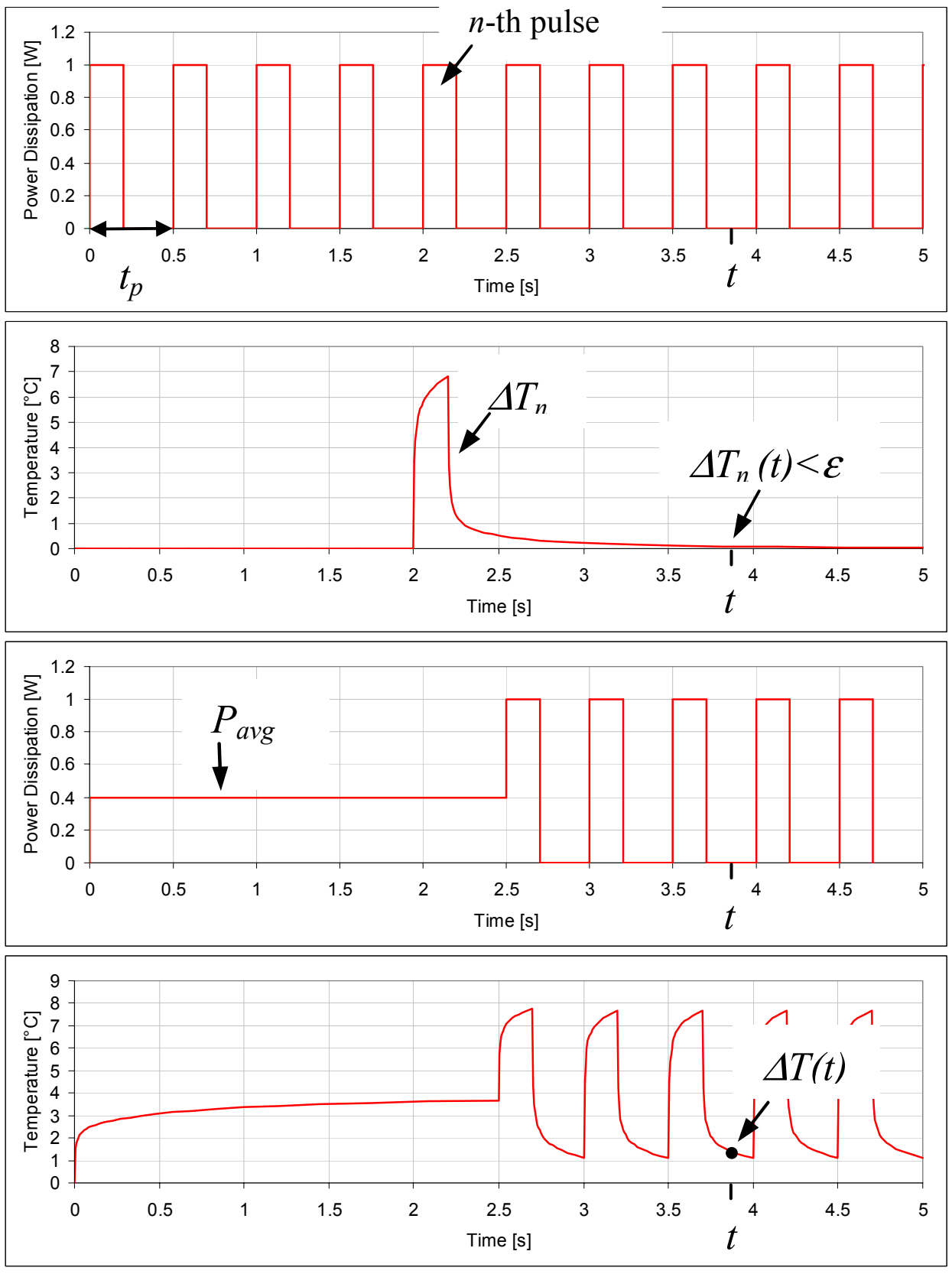

**Figure 6: The temperature response to the *n*-th pulse is < ε at time *t*; therefore the *n* first pulses are replaced by a pulse with constant power $P_{avg}$. The last plot shows the resulting temperature response, yielding the result at time *t*.**

Therefore the computing time can be reduced drastically if the first $n$ pulses of the periodic sequence are replaced by a single constant pulse of length $n \cdot t_p$ and average power $P_{avg} = E_{Pulse} / t_p$ , $E_{Pulse}$ being the energy per pulse. In practice, $n$ is chosen such that the temperature response $\Delta T_n(t)$ to the $n$-th pulse is < ε (small) at the time $t$ of interest and ≥ ε for all subsequent pulses > $n$ (figure 6).

## 5. APPLICATION TO AN AUTOMOTIVE ENGINE CONTROL DEVICE

The Infineon device TLE 6220 in a P-DSO-12 package controls the ignition of a four-cylinder engine. The power is dissipated periodically in four regions on the chip, named channel 1-4 (figure 7). The exposed pad of the package is soldered to a PCB, the bottom side of which being fixed at a temperature of 125°C.

The methods described herein shall be used to compute for this device the transient temperature response to a sequence of application specific pulses. Due to the symmetry of the device, all necessary (transfer) Zth-functions can be obtained by one transient simulation, heating only channel 1 and monitoring the temperature rise at the center of channels 1-4.

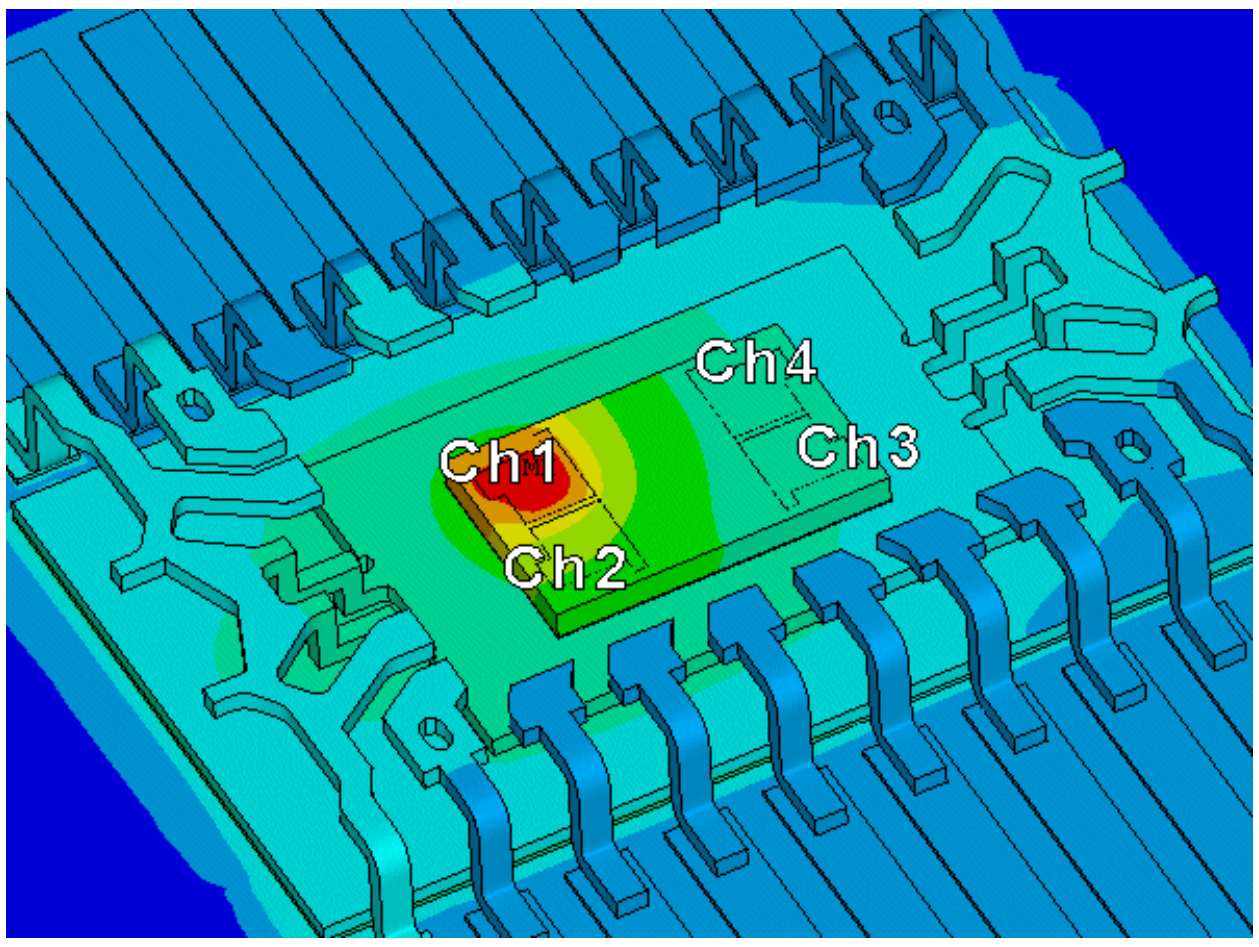

**Figure 7: Finite Element model of the TLE 6220**

Using the macro programming language of the Finite Element software ANSYS (APDL), the simulated step responses are written directly to a file which can be imported by a simulator without further post-processing. The standardized XML file format has been chosen for this purpose since it can be easily extended and XML parser exist for almost all programming languages.

The Zth functions have been computed for 74 equidistant points on a logarithmic timescale, ranging in between 0.5 µs and 1000 s (steady-state). Figure 8 shows the power profile which is to be applied to the channels: after a switch-on phase of 0.5 ms during which the power rises linearly to 1.0 W, the power remains constant until the channel is switched off after 4 ms. The energy stored in a connected load inductance is released within 0.3 ms, causing a peak power dissipation of 60 W in the device. This power profile is subsequently applied to all channels each shifted by 4 ms.

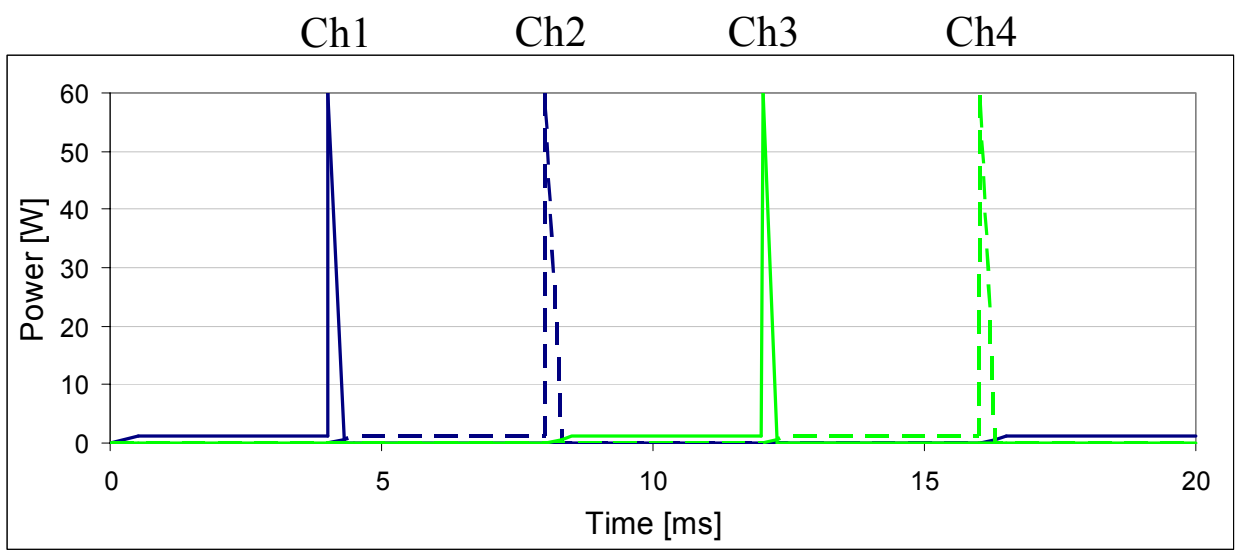

**Figure 8: Power profile**

In a direct transient FE simulation 544 points of time were necessary to resolve the temperature response in a time range from 0 to 20 ms. The simulation took about 11 h on a 2.8 GHz CPU compared to 90 min for the computation of the Zth-functions (74 points). Using these Zth-functions, an EXCEL simulator which implements the algorithm described above can compute the same temperature response within seconds (figure 9). In addition, changes to the power profile can be analyzed immediately, which would otherwise require a new time-consuming FE simulation.

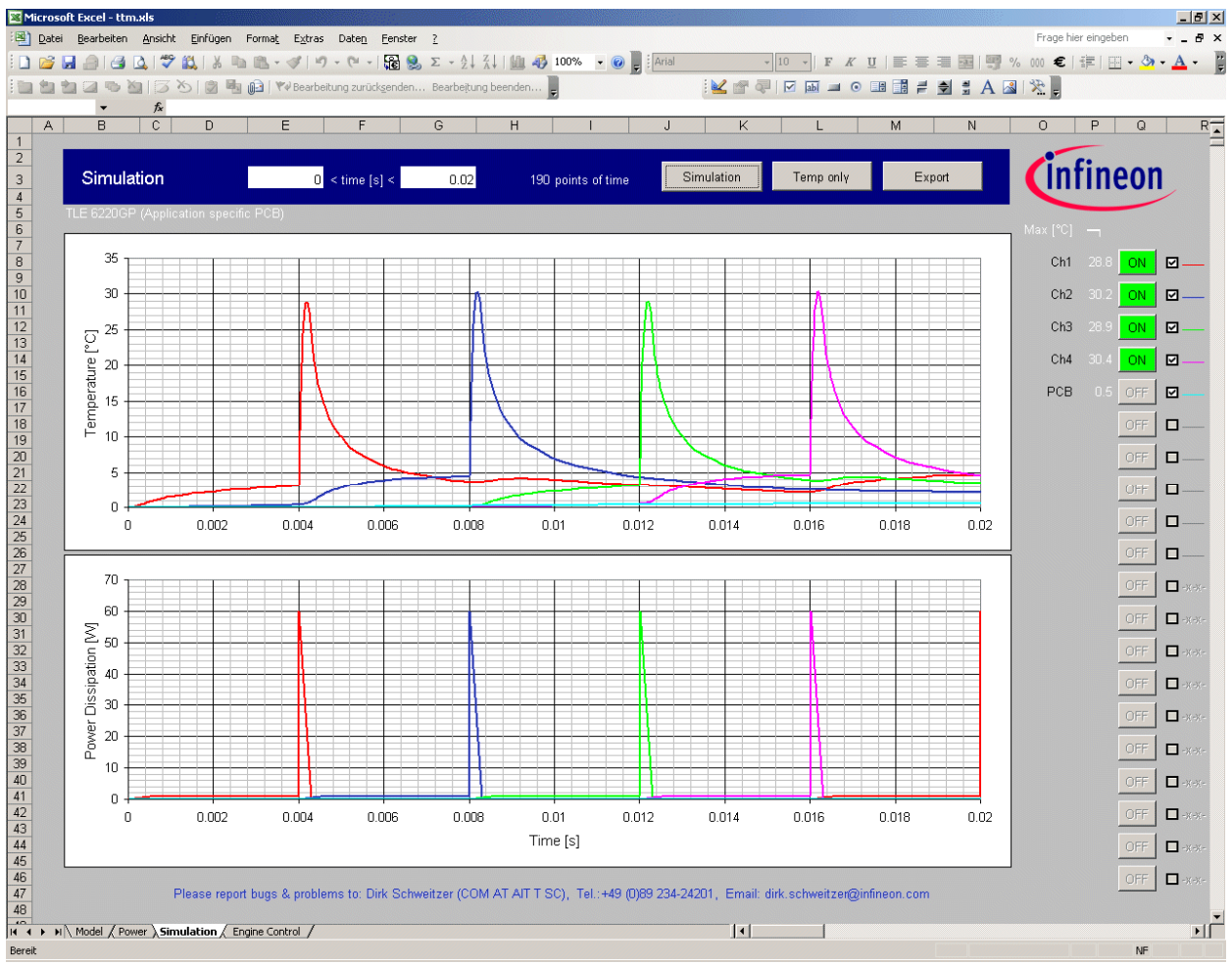

**Figure 9: EXCEL Thermal Transient Simulator**

In Figure 10 the temperature curves obtained by FE simulation and by the EXCEL simulator are so close, that differences are not distinguishable. The absolute error plot shows that the maximum difference < 0.3°C occurs during the short high power pulses. Discrepancies arise from numerical errors and non-linearities of the FE model (thermal conductivity of silicon).

## 6. CONCLUSION

Using interpolated Zth functions for the evaluation of the convolution integral in eq. (2), simulated or measured Zth curves can be used directly for the calculation of the temperature response to arbitrary power profiles in multi-source devices. This can be several heat-sources on a single chip as well as multi-chip devices or even several devices on a PCB. An implementation of the algorithm in EXCEL has demonstrated that the temperature curves for the selected source nodes of the model can be computed much faster than by a pure Finite Element simulation, even with the interpreted VBA programming language of EXCEL. This enables a fast analysis of different power profiles as well as the simulation of fast switching processes over a long period of time. A compiled code would perform even better, allowing also the simulation of temperature dependent power sources. The approximation error of the cubic spline fit is very small; deviations are mainly caused by non-linearities of the FE model.

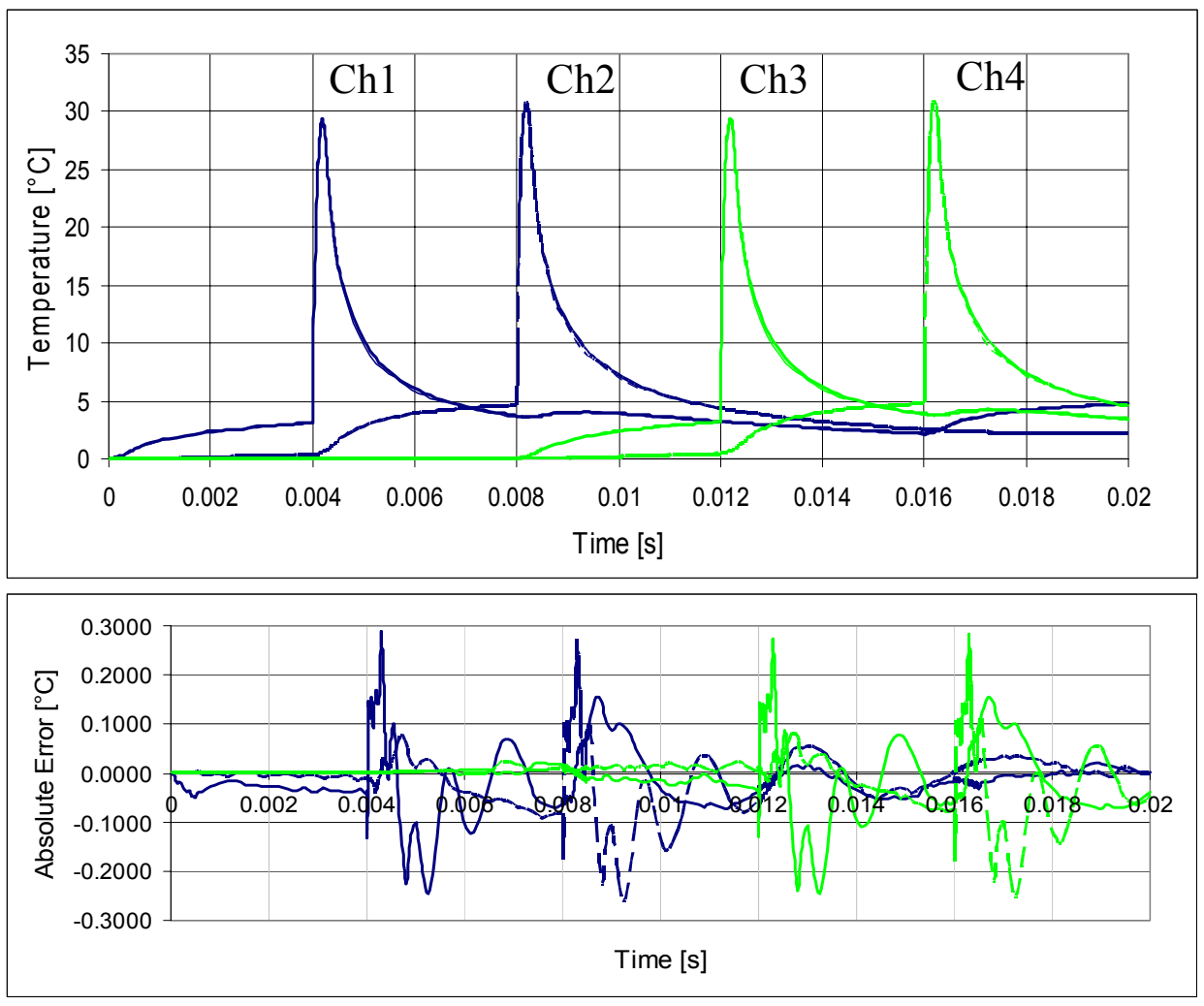

**Figure 10: Comparison FE simulation – EXCEL simulator**

## 7. REFERENCES

[1] Y.C. Gerstenmair and G. Wachutka, A New Procedure for the Calculation of the Temperature Development in Electronic Systems, *EPE'99 conference*, Lausanne, Switzerland, 1999.

[2] Y.C. Gerstenmair and G. Wachutka, Calculation of the Temperature Development in Electronic Systems by Convolution Integrals, *Proc.16th SEMITHERM*, San Jose, pp. 50-59, 2000.

[3] Vladimir Szekely, Identification of RC Networks by Deconvolution: Chances and Limits, *IEEE Trans. On Circuits and Systems vol. 45*, pp. 244-258, 1998.

[4] Torsten Hauck, and Tina Bohm, Thermal RC-Network Approach To Analyze Multichip Power Packages, *Proc. 16th SEMITHERM, San Jose,* pp. 227-234, 2000.

[5] W.H. Press, S.A. Teukolsky, W.T. Vetterling, and B.P. Flannery, *Numerical Recipes in C: The Art of Scientific Computing*, 2nd ed., Cambridge, Cambridge University Press, 1992.